\documentclass[letterpaper]{article}
\usepackage{amsmath}

\def\art#1{[\ref{#1}]}

\begin{document}

\title{\large{On the ``Application of the Kolmogorov-Smirnov test to CMB data:
Is the universe really weakly random?'', by Sigurd~K.~N\ae ss}}
\author{A.A.~Kocharyan\footnote{armen.kocharyan@monash.edu}\\
{\small School of Mathematical Sciences, Monash University, Clayton, Australia}}

\maketitle

\begin{center}
{\bf Abstract}
\end{center}

This short note is concerned with a recent paper by N\ae ss. We explain why the statements in the paper are absolutely irrelevant.

\vspace{0.3in}

Arnold in \art{Arnold_KSP}, \art{Arnold_UMN}, \art{Arnold_MMS}, \art{Arnold_FA} applies the stochastic parameter and the statistic introduced by Kolmogorov \art{K} to measure the objective stochasticity degree of datasets. He proves that Kolmogorov Stochasticity Parameter (KSP) method is mathematically sound, non-trivial, and universal  (cf. \art{AAK}).

KSP \art{GK_KSP} is applied to quantify the degree of randomness (stochasticity) in the temperature maps of the Cosmic Microwave Background radiation maps.

In a recent paper \art{SKN}, the author fails to understand the difference between KSP and Kolmogorov-Smirnov test (K-S test). A straw man argument is claimed against the results obtained in \art{WR}. It is stated that Kolmogorov-Smirnov test is applied to CMB data ``Application of the {\it Kolmogorov-Smirnov test} to CMB data.'' In reality, the papers \art{GK_KSP}, \art{WR} concern ``{\it Kolmogorov stochasticity parameter} measuring the randomness in the cosmic microwave background.'' Most importantly, it is obvious from Arnold's work that KSP is applicable to {\bf strongly correlated} datasets, too. 

One must read Arnold's papers to understand the key differences between KSP and K-S test. The very fact that Arnold in his paper series devoted to KSP does not mention K-S test at all, should be enough to conclude that the methods have different objectives. We hope our colleagues will consult literature before writing another irrelevant, critical paper (cf. \art{EW}).

\end{document}